\documentclass[a4paper,11pt]{article}
\usepackage[utf8]{inputenc}

\usepackage{amsfonts}
\usepackage{amssymb}
\usepackage{graphicx}
\usepackage{amsmath}
\usepackage{enumerate}
\usepackage{color}
\usepackage{cite}
 \usepackage{multirow}
\usepackage{float}

\RequirePackage{mathrsfs} \RequirePackage[sc]{mathpazo}
\RequirePackage{wasysym} \RequirePackage{setspace}\textheight=650pt
\title{ {\bf Thermodynamics of Charged AdS Black Holes in Extended Phases Space via M2-branes Background}}
\author{ M. Chabab$^1$, H. EL Moumni$^{1,2}$, K. Masmar$^1$\\
{\small $^{1}$High Energy Physics and Astrophysics Laboratory, FSSM,
 \small Cadi Ayyad University, Marrakesh, Morocco.
} \\
{\small $^{2}$  D\'{e}partement de Physique, Facult\'{e} des
Sciences, Universit\'{e} Ibn Zohr,
 Agadir, Morocco.} 
\\
 }
\date{\today}

\begin{document} \maketitle
\begin{abstract}
Motivated by a  recent work on asymptotically Ad$S_4$
black holes in M-theory, we investigate both thermodynamics and
thermodynamical geometry of Raissner-Nordstrom-AdS black holes from M2-branes.
More precisely, we study AdS black holes in $AdS_{4}\times
S^{7}$, with the number of M2-branes interpreted as a thermodynamical variable.  In this context, we calculate various thermodynamical quantities
including the chemical potential, and examine their
phase transitions along with  the corresponding stability behaviors. In addition, we also evaluate the thermodynamical curvatures of the  Weinhold, Ruppeiner and Quevedo metrics for M2-branes geometry to study the stability of such black
object.  We show that the singularities of these scalar curvature's metrics  reproduce similar stability results obtained by the phase transition program via the heat capacities in different ensembles either when the number of the M2 branes or the charge are held fixed. Also, we note that all results derived in  \cite{ana} are recovered in the limit of the vanishing charge. 
\end{abstract}
\newpage
\section{Introduction}

A more increasing interest has been recently devoted to the black hole physics and the connection with both string theory and thermodynamical models. Particularly,  many studies focus on the relationship between the gravity theories and the thermodynamical physics using
Anti-De Sitter geometries.  In this context, the laws of thermodynamics have been translated into laws of black holes  \cite{30,witten,4,5,50}. Hence, the phase transitions along with various   critical phenomena for AdS black holes have been extensively analysed in the framework of
different approaches \cite{6,a1,Dolan,Cliff,15Dolan}. Also, the equations of state describing   rotating black holes have
been interpreted by confronting them to  some known thermodynamical  ones, as Van der Waals gas \cite{our,our1,our2,hasan,KM,ourx,oury,Zhao}.  Emphasis has also been put on the free energy ands its behavior in the fixed charge ensemble. These studies shed some light on the thermodynamical criticality,  free energy, first order
phase transition and on understanding  of the behaviors near the critical points  with respect to 
the liquid-gas systems.

In this context, very recently the
thermodynamics and thermodynamical geometry for five dimensional
AdS black hole in type IIB superstring background known by
$AdS_5\times S^5$ \cite{Zhang,Zhang1, jhep45} have been scrutinezed. 
this geometry has been studied in many  places in connection with
AdS/CFT correspondence provides a very useful framework to investigate such geometry via the   equivalence between gravitational theories  in d-dimensional AdS space and
the conformal field theories (CFT) in a (d-1)-dimensional boundary of
such  AdS spaces \cite{Maldacena:1997re,ads,ads1,ads2}.  The number of colors has been interpreted  as  a thermodynamical variable in these works. In this respect, various thermodynamical quantities have been computed and the stability problem of
$AdS_5\times S^5$ black holes analysed by identifying the cosmological constant in
the bulk with the number of colors.

All these recent inspiring works  on asymptotically
Ad$S_4$ black holes in M-theory \cite{m0,m1,m2,m3} motivate us to 
study the thermodynamics and thermodynamical geometry of $AdS_{4}\times
S^{7}$, from the physics of M2-branes, where 
we interpret the number of M2 as a thermodynamical
variable as in \cite{ana}.  We then discuss the stability of such solutions and examine
the  corresponding first phase transition by analysing  various 
relevant quantities including  the chemical potential, free energy and heat capacitites.  Besides, we also
evaluate the thermodynamical curvatures from the 
Weinhold, Ruppeiner and Quevedo metrics for M2-branes geometry 
and study the corresponding stability problems via their singularities.
\\
The paper is arranged as follows:    In section 2 we discuss  thermodynamic
properties and stability of  the charged  black holes in  $AdS_{4}\times
S^{7}$,  by assuming the number of M2-branes
as a thermodynamical variable.  Section 3  and 4 are  devoted to show that similar results are recovered  through thermodynamical curvature calculations associated with the Weinhold, Ruppeiner and  Quevedo metrics. Our conclusion is drawn in section 5.

\section{ Thermodynamics of black holes in  $AdS_4\times S^7$  space}

In this section, we investigate  the phase transition of  the Reissner Nordstrom-AdS
black holes in M-theory in the presence of solitonic objects. Here we recall that, at low energy, M-theory describes  an eleven
dimensional supergravity. This theory, as proposed by Witten ,
can produce some nonperturbative limits  of superstring models
after its compactification  on  particular geometries \cite{Witten:1997sc}.

 First, let us consider the case of  M2-brane.  The corresponding geometry
is $AdS_4\times S^7$. In such a geometric background, the line
element of the black M2-brane metric  is given by \cite{a58,a}
\begin{equation}\label{ds4}
ds^2= \frac{r^4}{L^4}\left(-f dt ^2+\sum_{i=1}^2 dx_i^2 \right)+\frac{L^2}{r^2}f^{-1}dr^2+ L^2 d\Omega_7^2,
\end{equation}
where $d\Omega_7^2$ is the metric of seven-dimensional sphere with
unit radius.  In this solution, the metric function reads as follows
\begin{equation}\label{f4}
f=1-\frac{m}{r}+\frac{q^2}{r^2}+\frac{r^2}{L^2},
\end{equation}
where $L$ is the AdS radius and $m$ and $q$  are integration constants. The
cosmological constant is $\Lambda = -6/L^2$. Form M-theory point of
view,  the  eleven-dimensional spacetime  in Eq.(\ref{ds4}) can be
interpreted  as the near horizon geometry of $N$ coincident
configurations of  M2-branes. In  this background, the AdS radius
$L$ is linked to the
 M2-brane number $N$   via the relation  \cite{a,a7,ana}
\begin{equation}
L^9={N^3/2} \frac{\kappa_{11}^2 \sqrt{2}}{ \pi^5}.
\end{equation}
According to the proposition reported in \cite{Zhang,Zhang1,jhep45,ana},
we consider  the cosmological constant as the number of
M2-branes in the M theory background and its conjugate quantity as
the associated chemical potential.

The event horizon $r_h$ of the corresponding  black hole is
determined by solving  the equation $f = 0$.   From 
Eq.(\ref{f4}), the mass  of the black hole can be  written  as
\begin{equation}
M=\frac{m \omega_2}{8 \pi  G_4}=\frac{r \omega_2 \left(L^2+r^2\right)}{8 \pi  G_4 L^2}+\frac{2 \pi  G_4 Q^2}{r \omega_2}.\footnote{where $\omega_d=\frac{2 \pi ^{\frac{d+1}{2}}}{\Gamma
\left(\frac{d+1}{2}\right)}$.}
\end{equation}

where the charge of the black hole $Q$ is related to the constant $q$ through the formula,
\begin{equation}
Q=\frac{\omega_2}{4\pi G_4}q.
\end{equation}

 The Bekenstein-Hawking entropy formula
of the black hole reads as,
\begin{equation}
S=\left.\frac{A}{4 G_4}\right.=\frac{\omega_2  r^2}{4 G_4}.
\end{equation}
 Here we recall that  four-dimensional Newton gravitational constant is
related to the eleven-dimensional one as 
\begin{equation}
G_4=\frac{3 G_{11}
}{2 \pi  \omega_{2}L^{4}}.\, 
\end{equation}
For simplicity reason,  we  use 
$G_{11}=\kappa_{11}=1$ in the remainder of  the paper. In this way,  the black hole mass can
be expressed  as a function of $N$ and $S$,
\begin{equation}\label{M4}
M(S,N)=\frac{3 \sqrt[9]{2} \pi ^{11/9} \sqrt[3]{N} Q^2+3 \sqrt[3]{\pi } S^2+8
   \sqrt[3]{2} N S}{4\ 2^{13/18} \sqrt{3} \pi ^{11/18} N^{2/3}
   \sqrt{S}}
\end{equation}
Using  the standard thermodynamic relation $dM = TdS + \mu\ dN+ \Phi \ dQ$, the
corresponding  temperature takes the following form
\begin{equation}\label{T4}
T=\left.\frac{\partial M(S,N)}{\partial S}\right|_N=\frac{-3 \sqrt[9]{2} \pi ^{11/9} \sqrt[3]{N} Q^2+9 \sqrt[3]{\pi } S^2+8
   \sqrt[3]{2} N S}{8\ 2^{13/18} \sqrt{3} \pi ^{11/18} N^{2/3}
   S^{3/2}}.
\end{equation}
This quantity  can be identified with  the Hawking temperature of
the black hole.  Using eq. (\ref{M4}) the chemical potential $\mu$
conjugate to the number of  M2-branes is given by
\begin{equation}\label{mu4}
\mu=\left.\frac{\partial M_4(S,N)}{\partial N}\right|_S=\frac{-3 \sqrt[9]{2} \pi ^{11/9} \sqrt[3]{N} Q^2-6 \sqrt[3]{\pi } S^2+8
   \sqrt[3]{2} N S}{12\ 2^{13/18} \sqrt{3} \pi ^{11/18} N^{5/3}
   \sqrt{S}}.
\end{equation}

It defines the measure of the energy cost to the system when one
increases the variable  $N$.

while the electric potential reads as

\begin{equation}
\Phi=\left.\frac{\partial M(S,N)}{\partial Q}\right|_S=\frac{\sqrt{3} \pi ^{11/18} Q}{2\ 2^{11/18} \sqrt[3]{N} \sqrt{S}}.
\end{equation}

In terms of these quantities, the Helmholtz
 free energy  is expressed by,
\begin{equation}\label{ 4}
\mathcal{F}(T,N)=M-T\  S=\frac{9 \sqrt[9]{2} \pi ^{11/9} \sqrt[3]{N} Q^2-3 \sqrt[3]{\pi } S^2+8
   \sqrt[3]{2} N S}{8\ 2^{13/18} \sqrt{3} \pi ^{11/18} N^{2/3}
   \sqrt{S}}.
\end{equation}
Having calculated the relevant  thermodynamical quantities, we turn now to the
analysis of the corresponding  phase transition. For this, we study
the variation of the Hawking temperature as a function of the
entropy. 

\begin{figure}[!ht]
\begin{center}
\includegraphics[scale=1.]{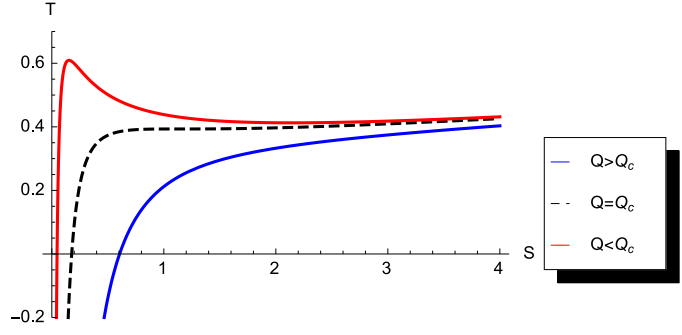}
\end{center}
 \caption{The temperature as function of the entropy $S$, with $N=3$} \vspace*{-.2cm}
 \label{fig1}
\end{figure}

This variation plotted in figure \ref{fig1} shows that Hawking temperature is
a monotonic function if $Q>Q_c$, but when $Q\leq Q_c$, it presents a critical point to be  determined by solving the system,

\begin{equation}
\left(\frac{\partial  T}{\partial S}\right)_{Q_c}=\left(\frac{\partial ^2 T}{\partial S^2}\right)_{Q_c}=0
\end{equation}

The solution of this equation is easily derived, 

\begin{equation}\label{critic}
Q_c=\frac{4\ 2^{5/18} N^{5/6}}{9 \pi ^{7/9}},\quad S_c=\frac{4 }{9}\sqrt[3]{\frac{2}{\pi }} N.
\end{equation}

\begin{figure}[!ht]
\begin{center}
\includegraphics[scale=1.]{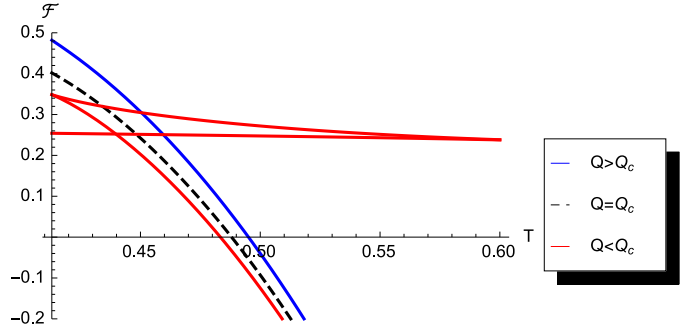}
\end{center}
 \caption{The   free energy  as function of the Temperature $T$.} \vspace*{-.2cm}
 \label{fig2}
\end{figure}

In figure  \ref{fig2}, we illustrate the  Helmholtz free energy  as
function of  the Hawking temperature $T$ for some fixed values of
$N$.

The sign change of the free energy indicates Hawking-Page phase transition, which occurs at
\begin{equation}
S_{HP}=\frac{\sqrt[18]{2} \left(4\ 2^{5/18} N+\sqrt{16\ 2^{5/9} N^2+27 \pi
   ^{14/9} \sqrt[3]{N} Q^2}\right)}{3 \sqrt[3]{\pi }}.
\end{equation}
It can be seen  that the "swallow tail", a type signal for the first phase transition, between small black hole and large one.

\begin{figure}[!ht]
\begin{center}
\includegraphics[scale=1.]{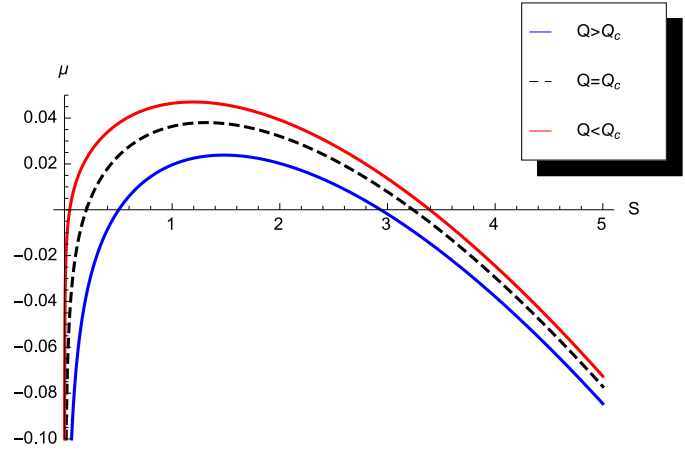}
\end{center}
 \caption{The chemical potential $\mu$ as function of the entropy for  $N=3$.} \vspace*{-.2cm}
 \label{fig3}
\end{figure}

To study the phase transition,  we vary the chemical potential in
terms of the entropy, and plot in figure \ref{fig3}  such a
variation for a   fixed value  of  $N$.

From the figure we can see that he chemical potential becomes positive when the entropy lies within the interval $S_-\leq S\leq S_+$ with
\begin{equation}
S_\pm=\frac{4 \sqrt[3]{2} N\pm 2^{5/9} \sqrt{8\ 2^{5/9} N^2-9 \pi ^{14/9}
   \sqrt[3]{N} Q^2}}{6 \sqrt[3]{\pi }}
\end{equation}

%

\begin{figure}[!ht]
\begin{center}
\includegraphics[scale=1.]{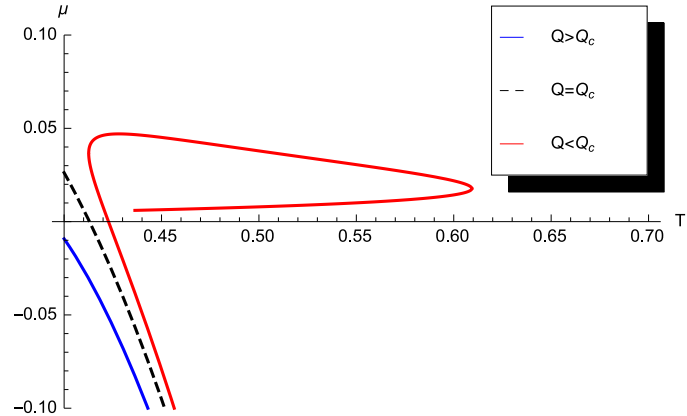}
\end{center}
 \caption{The chemical potential $\mu$ as function of the temperature $T$, with $N=3$.} \vspace*{-.2cm}
 \label{fig4}
\end{figure}

Furthermore, we also plot in figure \ref{fig4} the behavior of  the chemical potential as a function of temperature $T$ for a fixed $N$ .
 
From figure \ref{fig4} we can see that there exists a multivalued region, which just corresponds to the unstable region of the black hole with a negative heat capacity (red line in figure \ref{fig1}).


To illustrate the effect  of the number of the M2-branes, we discuss the
behavior of the chemical potential $\mu$ in terms of such a
variable as shown in figure \ref{fig5}.
\begin{figure}[!ht]
\begin{center}
\includegraphics[scale=1.]{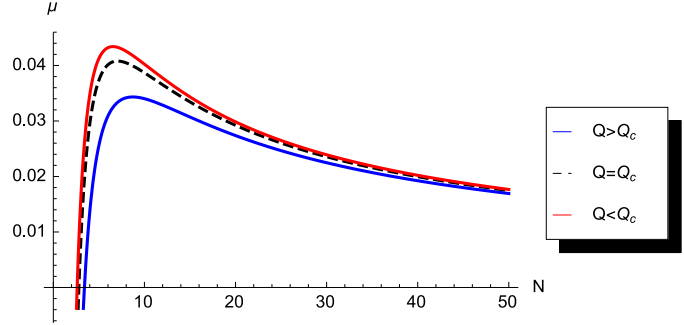}
\end{center}
 \caption{The chemical potential $\mu$ as function of $N$, we have set $S_4=4$.} \vspace*{-.2cm}
 \label{fig5}
\end{figure}

We clearly see that the chemical potential $\mu$ presents a maximum at
\begin{equation}
N_{max}=\frac{3 \sqrt[3]{\pi } \left(\frac{2 \pi ^2 Q^6}{\sqrt[3]{15 Q^6 S^{5/2}+\sqrt{225 Q^{12}
   S^5-4 \pi ^3 Q^{18}}}}+\sqrt[3]{2} \pi  \sqrt[3]{15 Q^6 S^{5/2}+\sqrt{225 Q^{12} S^5-4
   \pi ^3 Q^{18}}}+5\ 2^{2/3} S^{5/2}\right)}{16 S^{3/2}}
\end{equation}
 We remark  that  this  is quite different from the classical gas having a  negative chemical potential. In the case where  the
chemical potential approaches to zero and becomes positive, quantum
effects should be considered and become relevant in the discussion \cite{jhep45}.

In the subsequent sections, we consider thermodynamical geometry
 of the M2-branes black holes in the extended phase space
 and study the stability  problem when either $N$  or the charge is held fixed.

\section{Geothermodynamics  and phase transition of charged AdS  black holes  with fixed $N$ case}
Here we discuss the  geothermodynamics  of the charged AdS black holes
in $AdS_{4}\times S^{7}$:  Our analysis will focus on the singular limits
of certain thermodynamical quantities, including the heat capacities and scalar curvatures, 
which are relevant in the study of  the stability of such
black hole solution.

To do this, the  number of branes $N$ should be held
fixed  to consider the thermodynamics in the canonical ensemble. For a fixed $N$, the heat
capacities  for M2-branes AdS black hole are  given respectively by,

\begin{equation}\label{eq18x}
C_{Q,N}=T\left(\frac{\partial S}{\partial T}\right)_{Q,N}= T_4\left( \frac{8 \sqrt[3]{2} N+18 \sqrt[3]{\pi } S}{-3 \sqrt[9]{2} \pi ^{11/9}
   \sqrt[3]{N} Q^2+9 \sqrt[3]{\pi } S^2+8 \sqrt[3]{2} N S}-\frac{3}{2
   S}\right)^{-1}
   \end{equation}
  
   \begin{equation}\label{eq19}
   C_{\Phi,N}=T\left(\frac{\partial S}{\partial T}\right)_{\Phi,N}= \left(\frac{9 S}{-3 \sqrt[9]{2} \pi ^{8/9} \sqrt[3]{N} Q^2+9 S^2+8
   \sqrt[3]{\frac{2}{\pi }} N S}-\frac{1}{2 S}\right)^{-1}
\end{equation}

\begin{figure}[!ht]
\begin{center}
\includegraphics[scale=1.]{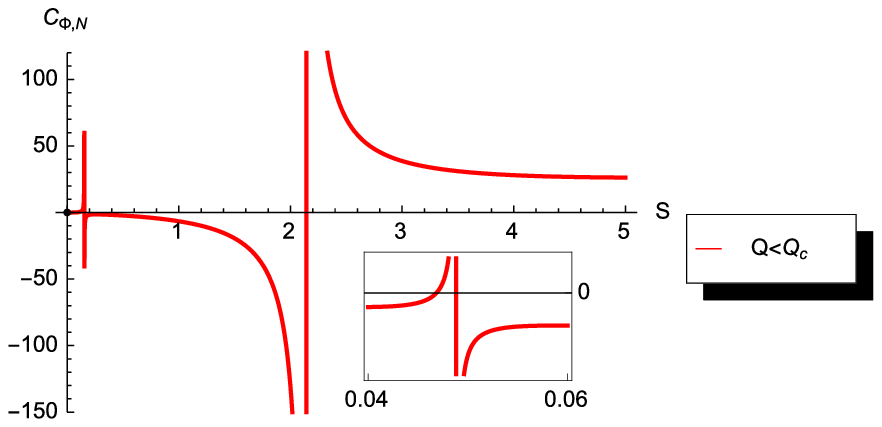}
\end{center}
\vspace*{-.2cm} \caption{ The heat capacity in the case with a fixed $N = 3$ as a function of entropy $S$ for  $Q=0.5\ Q_c$.}
\label{cnf}
 \label{fig5}
\end{figure}

In the canonical ensemble with fixed $N$, a critical point exists and is given by the Eq.(\ref{critic}). The behavior of the  $C_{\Phi,N}$ as function of the entropy is plotted in the figure \ref{cnf}.

 From the figure we can see that the $C_{\Phi,N }$ presents two singularities at 
\begin{equation}
S_{\Phi,\pm}=\frac{1}{9} \left(4 \sqrt[3]{\frac{2}{\pi }} N\pm \sqrt{16 \left(\frac{2}{\pi
   }\right)^{2/3} N^2-27 \sqrt[9]{2} \pi ^{8/9} \sqrt[3]{N}
   Q^2}\right)
\end{equation}

\begin{figure}[!ht]
\begin{center}
\includegraphics[scale=1.]{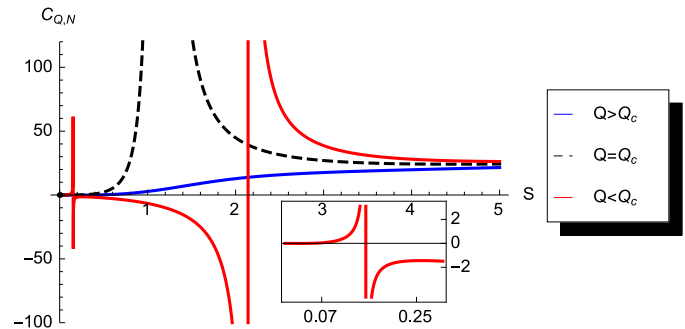}
\end{center}
\vspace*{-.2cm} \caption{ The heat capacity in the case with a fixed $N = 3$ as a function of entropy $S$.}
\label{cnq}
\end{figure}

The heat capacity $C_{Q,N}$ is plotted in figure \ref{cnq}

We see that it is consistent with the temperature shown in the figure \ref{fig1} (red line). For small  and large black hole the heat capacity is always positive, while for the intermediate black holes it is negative when $Q$ is less than the critical value, whereas it is always positive in the case when the charge is larger than the critical point.

The heat capacity $C_{Q,N}$ under the  critical case  has two singularities at
\begin{equation}\label{cq}
S_{Q.\pm}=\frac{8 \sqrt[3]{2} N\pm 2 \sqrt{16\ 2^{2/3} N^2-81 \sqrt[9]{2} \pi
   ^{14/9} \sqrt[3]{N} Q^2}}{18 \sqrt[3]{\pi }}
\end{equation}

these two singularities coincide for $Q=Q_c$, and (\ref{cq}) becomes 
\begin{equation}
S_Q=\frac{4}{9} \sqrt[3]{\frac{2}{\pi }} \left(\sqrt{N^2-3\ 3^{2/3}
   \sqrt[3]{N}}+N\right)
\end{equation}

 We turn now our attention to the thermodynamical geometry  of the black hole to see whether the thermodynamical curvature  can reveal the singularities of these two specific heats.
The Weinhold metric \cite{z19} is defined as the second derivative of the internal energy with respect to the entropy and other extensive quantities in the energy representation, while the Ruppeiner metric \cite{z20} is related the Weinhold metric by a conformal factor of the temperature \cite{z22}.
 
 \begin{equation}\label{3.5}
 ds_R^2=\frac{1}{T} ds_W^2
 \end{equation}
 
 Notice that the Weinhold and Ruppeiner metrics, which depend on the choice of thermodynamic potentials, are not Legendre invariant. The Quevedo metric defined as \cite{hep54,hep55,hep56,hep57}
\begin{equation}\label{3.7}
g=\left(E^c\frac{\partial \phi}{\partial E^c}\right)\left(\eta_{ab}\delta_{bc}\frac{\partial^2 \phi}{\partial E^c \partial E^d}\right), \quad \eta_{cd}=diag(-1,1,\cdots,1)
\end{equation}
 
is a Legendre invariant. $\phi$ denotes the thermodynamic potential, $E^a$ and $I^a$ represent respectively the set of extensive variables and the set of the  intensive variable, while $a=1,2,\cdots, n$.
 
In this context we can evaluate the thermodynamical curvature of the black hole.  For the Weinhold metric, 
 
\begin{equation}\label{3.8}
g^W=\left(\begin{array}{cc}M_{SS} & M_{S Q} \\M_{Q S} & M_{Q Q}\end{array}\right),
\end{equation}
where $M_{ij}$ stands for $\partial^2 M/ \partial x^i\partial x^j$, and $x^1=S$, $x^2=Q$,  we can see that its scalar curvature is derived via a direct calculation, simply by substituting Eq. (\ref{M4})  into Eq. (\ref{3.8}), 

\begin{equation}
R_1^W=-\frac{64 \sqrt[18]{2} \sqrt{3} \pi ^{11/18} N^{5/3} S^{3/2}}{\left(3
   \sqrt[9]{2} \pi ^{11/9} \sqrt[3]{N} Q^2+9 \sqrt[3]{\pi } S^2-8 \sqrt[3]{2}
   N S\right)^2}
\end{equation}

While the Ruppeiner metric,  deduced from Eq.(\ref{3.5}), is given by

\begin{equation}\label{3.9}
g^R=\frac{1}{T}\left(\begin{array}{cc}M_{SS} & M_{SQ} \\M_{Q S} & M_{QQ}\end{array}\right),
\end{equation}

with the following curvature,

\begin{equation}
R_1^R=\frac{A_1}{B_2}
\end{equation}
where,
\begin{eqnarray}  \nonumber
A_1&=&162 \pi ^{2/3} S \left[-486\ 2^{4/9} \pi ^7 N^2 Q^{12}+6480\ 2^{2/3} \pi
   ^{52/9} N^{8/3} Q^{10} S+1296\ 2^{2/9} \pi ^{2/3} N^{4/3} S^9\right. \\ \nonumber
  &\times& \left. \left(896\ 2^{5/9} N^{5/3}-351 \pi ^{14/9} Q^2\right)-432 \sqrt[3]{2} \pi
   ^{41/9} N^{5/3} Q^8 S^2 \left(80\ 2^{5/9} N^{5/3}-9 \pi ^{14/9}
   Q^2\right)\right.\\\nonumber
   &+&\left.144 \sqrt[9]{2} \pi ^{10/3} N^{7/3} Q^6 S^3 \left(1280
   N^{5/3}-261\ 2^{4/9} \pi ^{14/9} Q^2\right)+18 \sqrt[3]{\pi }
   N^{2/3} S^8 \right.\\\nonumber
   &\times&\left.
   \left(90112 \sqrt[9]{2} N^{10/3}+5103 \pi ^{28/9}
   Q^4-62208\ 2^{5/9} \pi ^{14/9} N^{5/3} Q^2\right)-96 \sqrt[3]{2}
   N^{5/3} S^7 \right.\\ \nonumber
   &\times&\left.
   \left(-4096 \sqrt[9]{2} N^{10/3}-3645 \pi ^{28/9}
   Q^4+8640\ 2^{5/9} \pi ^{14/9} N^{5/3} Q^2\right)+48 \sqrt[9]{2} \pi ^{11/9}
   N Q^2 S^6\right.\\ \nonumber
   &\times&\left. \left(-4096 \sqrt[9]{2} N^{10/3}-729 \pi ^{28/9}
   Q^4+5184\ 2^{5/9} \pi ^{14/9} N^{5/3} Q^2\right)-32 \pi ^{8/9} N^2
   Q^2 S^5 \right.\\ \nonumber
   &\times&\left.\left(-4096\ 2^{5/9} N^{10/3}+243\ 2^{4/9} \pi ^{28/9} Q^4+8064 \pi
   ^{14/9} N^{5/3} Q^2\right)+6\ 2^{2/9} \pi ^{19/9} N^{4/3} Q^4 S^4\right.\\ \nonumber
   &\times&\left.
   \left(-40960 \sqrt[9]{2} N^{10/3}-729 \pi ^{28/9} Q^4+20736\ 2^{5/9} \pi
   ^{14/9} N^{5/3} Q^2\right)-59049\ 2^{7/9} \pi ^{5/3} S^{12}\right.\\ 
   &-&\left.104976
   \sqrt[9]{2} \pi ^{4/3} N S^{11}+559872\ 2^{4/9} \pi  N^2
   S^{10}\right]\\
   B_1&=&\left(3 \sqrt[9]{2} \pi ^{11/9} \sqrt[3]{N} Q^2-9
   \sqrt[3]{\pi } S^2-8 \sqrt[3]{2} N S\right)^3 \\ \nonumber
   &\times&\left(-18 \sqrt[9]{2} \pi
   ^{22/9} N^{2/3} Q^4+96 \sqrt[3]{2} \pi ^{11/9} N^{4/3} Q^2 S+81\
   2^{8/9} \pi ^{2/3} S^4-128\ 2^{5/9} N^2 S^2\right)^2
   \end{eqnarray}

\begin{center}
\begin{figure}[!ht]
\begin{tabbing}
\hspace{9cm}\=\kill
\includegraphics[scale=1]{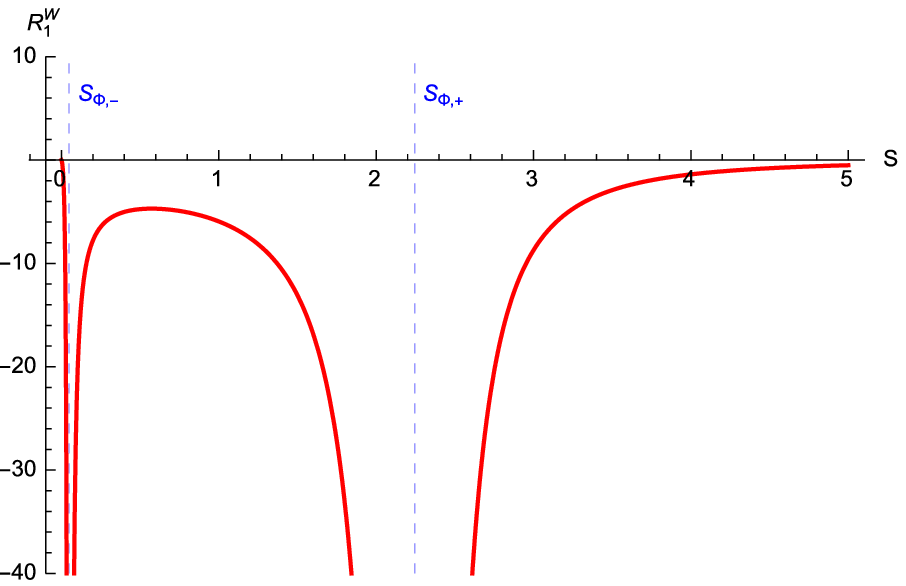}\>\includegraphics[scale=1]{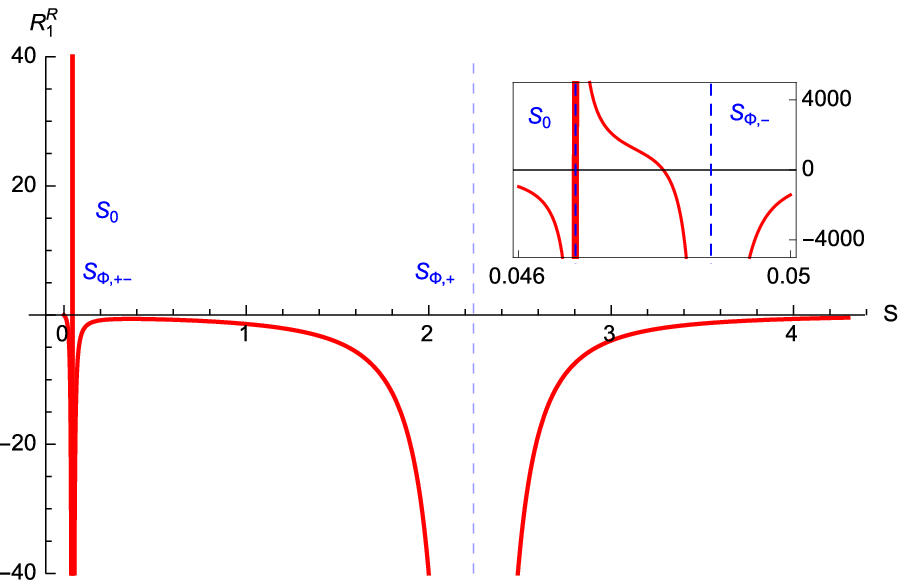} \\
\end{tabbing}
\vspace*{-.2cm} \caption{The scalar curvatures  of Weinhold and Ruppeiner  metrics vs entropy  with $N = 3$   and $Q=0.5\ Q_c$.}
\label{RWRR}
\end{figure}
\end{center}

In the figure \ref{RWRR} we plot the scalar curvatures of the Weinhold and Ruppeiner  metrics where the charge is less than critical one.

\newpage

From Fig.\ref{RWRR} we see that scalar curvature of Weinhold and Ruppeiner metrics reveal both  the same singularities $S_{\phi,\pm}$ of the heat capacity $C_{\Phi,N}$. The Ruppeiner metric presents a further singularity in $S_0$ where the black hole is extremal $T=0$. Hence both Weinhold and Ruppeiner metric are able to show phase transition of the black hole in the fixed $\Phi$ ensemble.

The Quevedo metric is  defined by,

\begin{equation}\label{3.9}
g^Q=(S\, T+ Q\, \Phi)\left(\begin{array}{cc}-M_{SS} & 0 \\0 & M_{Q Q}\end{array}\right),
\end{equation}

Using Eqs.(\ref{M4},\ref{T4}) and Eq.(\ref{3.9}) we show that the scalar curvature reads as, 
\begin{equation}
R_1^Q=\frac{A_2}{B_2}
\end{equation}
   with,
\begin{eqnarray}\nonumber
A_2&=&-768 \sqrt[9]{2} \pi ^{11/9} N^{4/3} S \left[4374 \sqrt[9]{2} \pi ^{44/9}
   N^{4/3} Q^8+131220 \pi ^4 N Q^6 S^2+19440 \sqrt[3]{2} \pi ^{11/3}
   N^2 Q^6 S\right.\\\nonumber
   &+&\left.104976\ 2^{8/9} \pi ^{28/9} N^{2/3} Q^4 S^4+85536\
   2^{2/9} \pi ^{25/9} N^{5/3} Q^4 S^3+17280\ 2^{5/9} \pi ^{22/9}
   N^{8/3} Q^4 S^2\right.\\\nonumber
   &+&\left.21870\ 2^{7/9} \pi ^{20/9} \sqrt[3]{N} Q^2
   S^6+19440 \sqrt[9]{2} \pi ^{17/9} N^{4/3} Q^2 S^5-34560\ 2^{4/9} \pi ^{14/9}
   N^{7/3} Q^2 S^4\right.\\\nonumber
   &+&\left.21504\ 2^{7/9} \pi ^{11/9} N^{10/3} Q^2 S^3-19683\
   2^{2/3} \pi ^{4/3} S^8-46656 \pi  N S^7-10368 \sqrt[3]{2} \pi ^{2/3}
   N^2 S^6\right.\\
   &+&\left.18432\ 2^{2/3} \sqrt[3]{\pi } N^3 S^5+32768 N^4
   S^4\right]\\ \nonumber
   B_2&=&\left(9\ 2^{7/9} \pi ^{11/9} \sqrt[3]{N} Q^2+9\ 2^{2/3}
   \sqrt[3]{\pi } S^2+16 N S\right)^2 \left(9 \sqrt[9]{2} \pi ^{11/9}
   \sqrt[3]{N} Q^2+9 \sqrt[3]{\pi } S^2-8 \sqrt[3]{2} N S\right)^2  \\ 
   &\times&
   \left(9 \sqrt[9]{2} \pi ^{11/9} \sqrt[3]{N} Q^2+9 \sqrt[3]{\pi } S^2+8
   \sqrt[3]{2} N S\right)^2
\end{eqnarray}

In the next figure, we plot $R_1^Q$ in terms of the entropy for a fixed $N$  (here $N=3$).

\begin{center}
\begin{figure}[!ht]
\begin{tabbing}
\hspace{9cm}\=\kill
\includegraphics[scale=1]{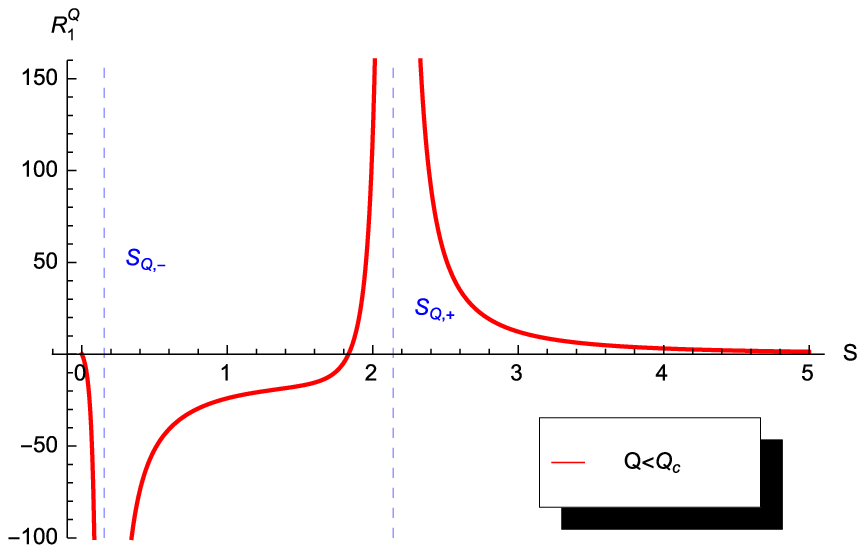}\>\includegraphics[scale=1]{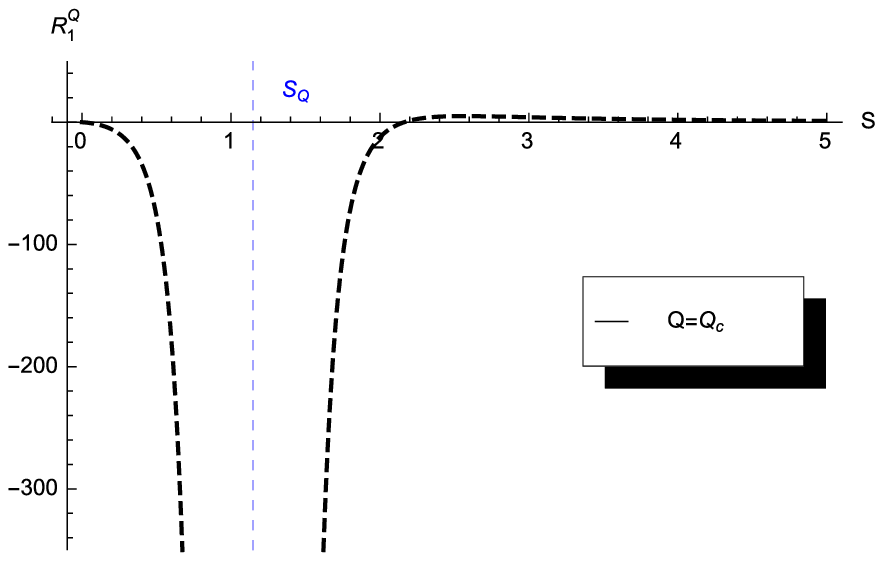} \\
\end{tabbing}
\vspace*{-.2cm} \caption{The scalar curvature vs entropy for the Quevedo metric case with $N = 3$ . Right side: $Q=Q_c$, Left side: $Q<Q_c$.}
\label{figrq}
\end{figure}
\end{center}

Under the critical scheme $Q<Q_c$ the scalar curvature of the Quevedo metric presents two singularities at $S_{Q,\pm}$ which are the same as those of the heat capacity  $C_{Q,N}$ shown in figure \ref{cnq} (red line). When $Q=Q_c$, the two  singularities $S_{Q,\pm}$ coincide to one $S_Q$ (represented by dashed black line). That mean that the Quevedo metric can reveal  the phase transition in the fixed charge ensemble.

\section{Geothermodynamics  and phase transition of charged AdS  black holes  with  fixed charge $Q$ }
In this section we study the thermodynamics geometry of the  $M2$-branes black holes in the canonical ensemble (fixed charge). That means that the charge of the black hole is not treated as thermodynamical variable but as a fixed external parameter.  The critical number of the M2-brane reads as,

\begin{equation} 
 N_c=\frac{9\ 3^{2/5} \pi ^{14/15} Q^{6/5}}{4\ 2^{11/15}}
 \end{equation}
 The heat capacity $C_{\mu,Q}$ with a fixed chemical potential is given by
\begin{equation}
C_{\mu,Q}=T\left(\frac{\partial T }{\partial S}\right)_{\mu, Q}^{-1}
\end{equation}

\begin{center}
\begin{figure}[!ht]
\begin{tabbing}
\hspace{9cm}\=\kill
\includegraphics[scale=1]{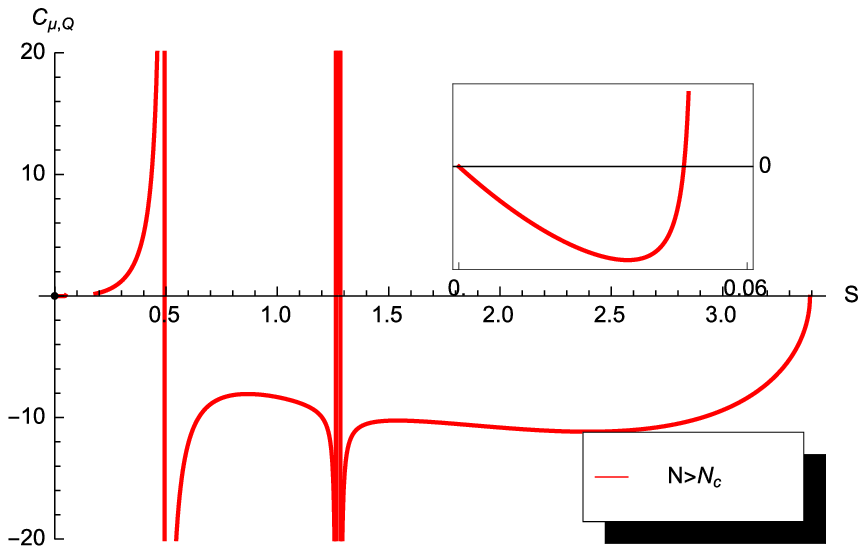}\>\includegraphics[scale=1]{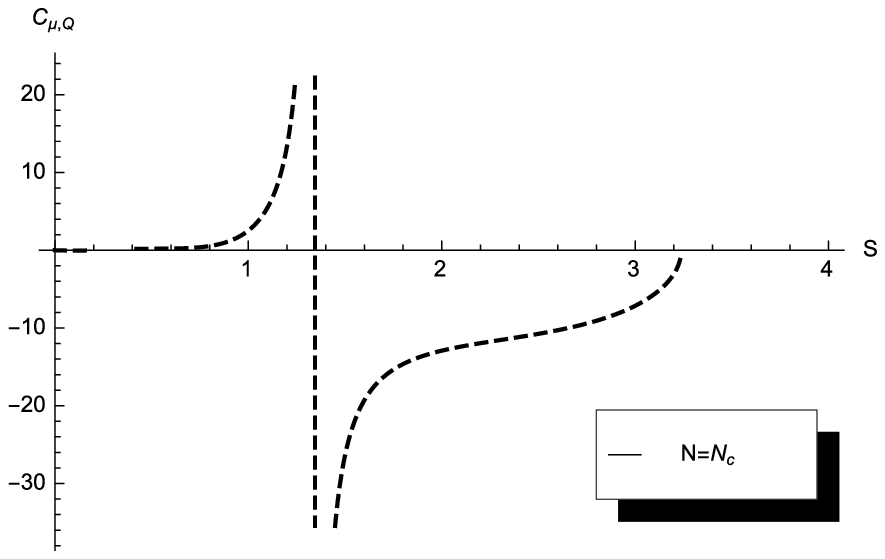} \\
\end{tabbing}
\vspace*{-.2cm} \caption{The specific heat $C_{\mu,Q}$ vs entropy with $Q=.55$, for $N>N_c$ and $N=N_c$.}
\label{cmuq}
\end{figure}
\end{center}

 The full expression of the $C_{\mu,Q}$, quite lengthy, is not given here. Instead we plot, in figure $\ref{cmuq}$, $C_{\mu,Q}$ in terms of the entropy in the critical sector.

From the left panel, we see that the heat capacity $C_{\mu,Q}$ presents two divergencies up to the critical regime, given numerically by $S_{\mu,-}\simeq0.5$ and $S_{\mu,+}\simeq1.27$ for $Q=0.55$. When $N=N_c$, these two singularities coincide, as shown in the right panel,  to only one singularity. 

In the fixed charge case the Weinhold metric can be expressed as

\begin{equation}\label{3.82}
g^W=\left(\begin{array}{cc}M_{SS} & M_{S N} \\M_{N S} & M_{N N}\end{array}\right),
\end{equation}

From a treatment similar to the calculation performed in the previous section, one  can derive  the full expression of the scalar curvatures of both the Weinhold and Ruppeiner metrics respectively. For the former one finds,

\begin{equation}
R_2^W= \frac{A_3}{B_3}
\end{equation}
with,
\begin{eqnarray} \nonumber
A_3&=&120\ 2^{5/6} \sqrt{3} \pi ^{11/6} N Q^2 S^{3/2} \left[-408\ 2^{4/9} \pi
   ^{11/9} N^{4/3} Q^2+837 \sqrt[9]{2} \pi ^{14/9} \sqrt[3]{N} Q^2
   S+1404 \pi ^{2/3} S^3\right.\\
   &+&\left.64 \sqrt[3]{2} N S \left(7 \sqrt[3]{2} N-27
   \sqrt[3]{\pi } S\right)\right]\\ \nonumber
   B_3&=&\left(-99\ 2^{2/9} \pi ^{22/9} N^{2/3}
   Q^4-486 \sqrt[9]{2} \pi ^{14/9} \sqrt[3]{N} Q^2 S^2+288\ 2^{4/9} \pi ^{11/9}
   N^{4/3} Q^2 S+54 \pi ^{2/3} S^4\right.\\
   &+&\left.96 \sqrt[3]{2 \pi } N S^3-64\
   2^{2/3} N^2 S^2\right)^2
\end{eqnarray}

while the result of the latter metric is, 

\begin{equation}
R_2^R=\frac{A_4}{B_4}
\end{equation}

with

\begin{eqnarray} \nonumber
A_4&=&15 \sqrt[3]{\pi } \sqrt[3]{N} \left[768 \sqrt[3]{\pi } N^{8/3}
   S \left(27 \pi ^3 Q^6-16 S^5\right)+99144 \sqrt[3]{2} \pi ^4 N^{2/3} Q^6
   S^3-432 (2 \pi )^{2/3} N^{5/3} S^2 \right.\\\nonumber
   &\times&\left.\left(261 \pi ^3 Q^6+8 S^5\right)+111537\
   2^{2/9} \pi ^{28/9} \sqrt[3]{N} Q^4 S^5-126360\ 2^{5/9} \pi ^{25/9}
   N^{4/3} Q^4 S^4\right.\\\nonumber
   &+&\left.123264\ 2^{8/9} \pi ^{22/9} N^{7/3} Q^4 S^3-55296\
   2^{2/9} \pi ^{19/9} N^{10/3} Q^4 S^2+99144 \sqrt[9]{2} \pi ^{20/9} Q^2
   S^7\right.\\\nonumber
   &-&\left.153600 \sqrt[9]{2} \pi ^{11/9} N^3 Q^2 S^4+49152\ 2^{4/9} \pi ^{8/9}
   N^4 Q^2 S^3-432\ 2^{7/9} \pi ^{14/9} N^2 Q^2 \left(3 \pi ^3
   Q^6+100 S^5\right)\right.\\
   &+&\left.81\ 2^{4/9} \pi ^{17/9} N Q^2 S \left(357 \pi ^3 Q^6+640
   S^5\right)+8192 \sqrt[3]{2} N^{11/3} S^5\right]\\ \nonumber
   B_4&=&\left(-3 \sqrt[9]{2} \pi
   ^{11/9} \sqrt[3]{N} Q^2+9 \sqrt[3]{\pi } S^2+8 \sqrt[3]{2} N
   S\right) \left(-99\ 2^{2/9} \pi ^{22/9} N^{2/3} Q^4-486 \sqrt[9]{2} \pi
   ^{14/9} \sqrt[3]{N} Q^2 S^2\right.\\
   &+&\left.288\ 2^{4/9} \pi ^{11/9} N^{4/3} Q^2
   S+54 \pi ^{2/3} S^4+96 \sqrt[3]{2 \pi } N S^3-64\ 2^{2/3} N^2
   S^2\right)^2
\end{eqnarray}


\begin{center}
\begin{figure}[!ht]
\begin{tabbing}
\hspace{9cm}\=\kill
\includegraphics[scale=1]{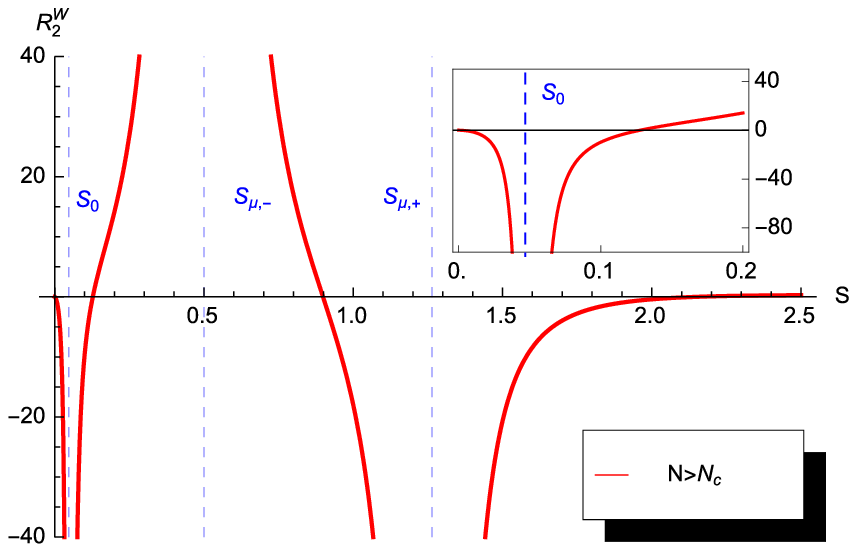}\>\includegraphics[scale=1]{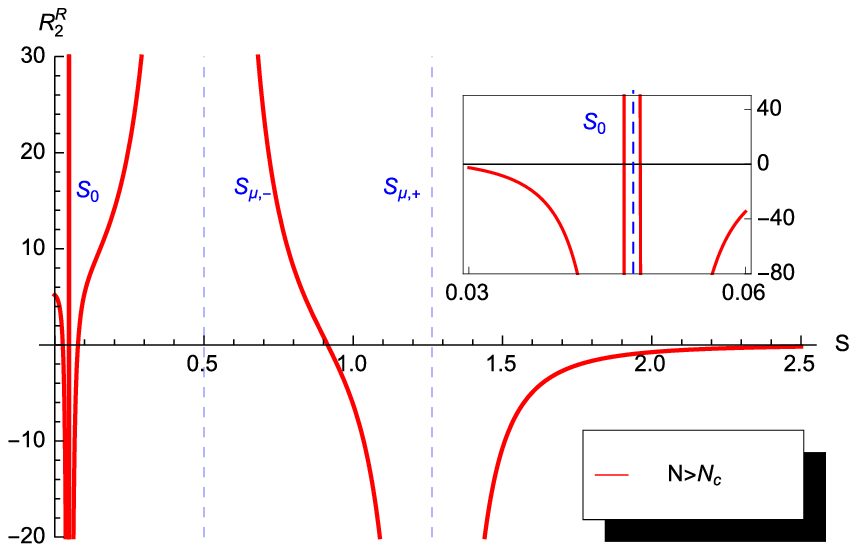} \\
\end{tabbing}
\vspace*{-.2cm} \caption{The scalar curvature of the Weinhold  and Ruppeiner metrics vs entropy for $N>N_c$, with $Q=0.55$.}
\label{rwrr}
\end{figure}
\end{center}

These two  scalar curvatures are plotted as functions of the entropy in figure \ref{rwrr}
which shows that the two metrics reproduce the results obtained in the previous section regarding the singularities of the heat capacity $C_{\mu,Q}$. Furthermore, the figure also shows the divergency of the scalar curvature of the Weinhold (left) and Ruppeiner (right) metrics when the entropy tends to the value $S_0$ for which the black hole becomes extremal $T=0$. 

Next we revisit the Quevedo' metric in the fixed charge case,  

\begin{equation}\label{3.92}
g^Q=(S\, T+ N\, \mu)\left(\begin{array}{cc}-M_{SS} & 0 \\0 & M_{N N}\end{array}\right),
\end{equation}
 and compute its corresponding scalar curvature given by,

\begin{equation}
R_2^Q=\frac{A_5}{B_5}
\end{equation}

The expression of $A_5$ and $B_5$ are found to be, 

\begin{eqnarray}\nonumber
A_5&=&864 \pi ^{14/9} N^{5/3} S^2 \left[768\ 2^{4/9} \sqrt[3]{\pi }
   N^{8/3} S \left(123 \pi ^3 Q^6-88 S^5\right)-243\ 2^{7/9} \pi 
   N^{2/3} S^3 \left(953 \pi ^3 Q^6-48 S^5\right)\right.\\\nonumber
   &-&\left.144 \sqrt[9]{2} \pi ^{2/3}
   N^{5/3} S^2 \left(5817 \pi ^3 Q^6+2176 S^5\right)-522450\ 2^{2/3} \pi
   ^{28/9} \sqrt[3]{N} Q^4 S^5+490320 \pi ^{25/9} N^{4/3} Q^4
   S^4\right.\\\nonumber
   &+&\left.2116224 \sqrt[3]{2} \pi ^{22/9} N^{7/3} Q^4 S^3-190464\ 2^{2/3} \pi
   ^{19/9} N^{10/3} Q^4 S^2-24786\ 2^{5/9} \pi ^{20/9} Q^2 S^7\right.\\\nonumber
   &-&\left.1588224\ 2^{5/9}
   \pi ^{11/9} N^3 Q^2 S^4+114688 (2 \pi )^{8/9} N^4 Q^2 S^3-864\
   2^{2/9} \pi ^{14/9} N^2 Q^2 \left(17 \pi ^3 Q^6-792 S^5\right)\right.\\
   &+&\left.81\ 2^{8/9}
   \pi ^{17/9} N Q^2 S \left(459 \pi ^3 Q^6+12520 S^5\right)+131072\ 2^{7/9}
   N^{11/3} S^5\right]\\
   B_5&=&5 \left(3 \sqrt[9]{2} \pi ^{11/9} \sqrt[3]{N}
   Q^2-3 \sqrt[3]{\pi } S^2-8 \sqrt[3]{2} N S\right)^3 \left(6 \sqrt[9]{2} \pi
   ^{11/9} \sqrt[3]{N} Q^2+15 \sqrt[3]{\pi } S^2-8 \sqrt[3]{2} N
   S\right)^2\\
   &\times& \left(9 \sqrt[9]{2} \pi ^{11/9} \sqrt[3]{N} Q^2+9 \sqrt[3]{\pi }
   S^2-8 \sqrt[3]{2} N S\right)^2
\end{eqnarray}

Illustration of $R_2^Q$ behaviour as a function of the entropy is seen in the next figure when $N>N_c$.

\begin{figure}[!ht]
\begin{center}
\includegraphics[scale=1.]{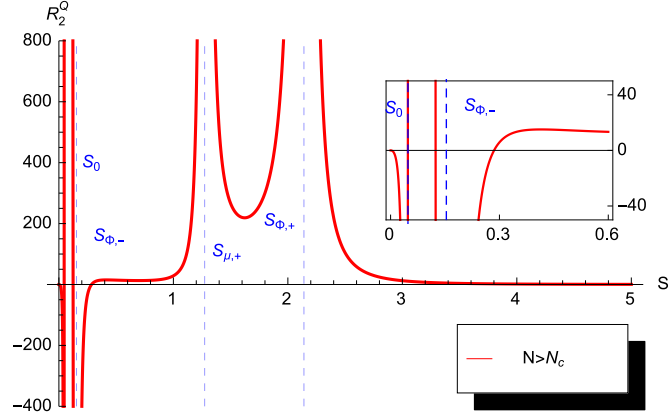}
\end{center}
\vspace*{-.2cm} \caption{ The heat capacity as a function of entropy $S$ for the two backgrounds in the case with a fixed $Q = 0.55$ .}
\label{rq2}
 \label{fig5}
\end{figure}

From the figure \ref{rq2} we can see that the Quevedo metric presents similar singularity's features,  here denoted by $S_{Q,\pm}$, as in the previous analysis of the  $C_{Q,N}$ for fixed $N$  shown in Eq.\ref{eq18x}. In addition we note that at $S=S_{\mu,+}$,  $C_{\mu,Q}$ becomes also singular, while at $S=S_0$ an additional singularity shows up signaling the extremal case. 

\section{Conclusion}
In this paper, we have explored  the thermodynamics and
thermodynamical geometry of  charged AdS black holes from M2-branes.
More concretely, by assuming the number of
M2-branes as a thermodynamical variable, we have considered AdS black holes in
$AdS_{4}\times S^{7}$. Then, we have
discussed the corresponding phase transition by computing the
relevant quantities. In particular, we have computed the
chemical potential and discussed the corresponding stabilities, the critical coordinates and the Helmoltez free energy.
 In addition,  we have also studied  the thermodynamical geometry
associated with   such  AdS black holes.  More precisely,
 we have derived the  scalar curvatures from the Weinhold, Ruppeiner  and Quevedo metrics and demonstrated that  these  thermodynamical properties are similar to those which show up in the phase transition program.  In the limit of the of the vanishing charge we recover all the results of \cite{ana}. We aim to extend this work to other geometries and black hole configurations.

\section*{Aknowledgements}
This work is supported in part by the GDRI project entitled: "Physique de l'infiniment petit et de l'infiniment grand" - P2IM  (France - Maroc).

\end{document}